\documentclass{aa}
\usepackage{graphicx}
\begin{document}
   \title{Near-Infrared photometry in the J, H and Kn bands for Polar Ring 
          Galaxies:}

   \subtitle{I. Data, structural parameters}

   \author{E. Iodice, 
          \inst{1,2}
          \and
           M. Arnaboldi
          \inst{1}
          \and 
           L.S. Sparke
          \inst{3}     
          \and 
           J.S. Gallagher
          \inst{3}     
          \and 
           K.C. Freeman
          \inst{4}     
          }

   \offprints{E. Iodice, \email{iodice@na.astro.it}}

   \institute{INAF - Osservatorio Astronomico di Capodimonte (OAC),\\ 
              via Moiariello 16, I-80131 Napoli
              \email{iodice@na.astro.it}
         \and
         International School for Advanced Studied (ISAS),
              via Beirut 2-4, I-34014 Trieste           
         \and   
          University of Wisconsin, Department of Astronomy,\\
             475 N. Charter St., Madison, WI
         \and  
          RSAA, Mt. Stromlo, Canberra, Cotter Road,\\ 
           Weston ACT 2611, Australia
             }

   \date{Received ; accepted }

   \abstract{
We present new Near-Infrared (NIR) observations, in the J, H and Kn bands,
for a sample of Polar Ring Galaxies (PRGs), selected from the 
Polar Ring Catalogue (Whitmore et al. \cite{PRC}).
Data were acquired with the CASPIR near-IR camera at the 2.3 m telescope 
of Mount Stromlo and Siding Spring Observatory. 
We report here on the detail morphological study for the central host galaxy 
and the polar structure in all PRGs of our sample. 
Total magnitudes, bulge-to-disk decomposition 
and structural parameters are computed for all objects.
These data are crucial for an accurate modeling of the stellar
population and the estimate of the star formation rates in the two
components.

   }

   \titlerunning{NIR Photometry for PRGs}
   \authorrunning{Iodice et al.}
   \maketitle
%

\section{Introduction}

Polar Ring Galaxies (PRGs) are classified as dynamically peculiar systems, 
as they show the coexistence of two luminous components, the central
host  galaxy and ring, with their angular momentum vectors in two nearly
orthogonal  planes (Schweizer et al. \cite{Schweizer83}; Whitmore et
al. \cite{PRC}).
The presence of almost two perpendicular angular momentum vectors 
cannot be explained through the collapse of a single 
protogalactic cloud: a ``second event'' must have occurred in the formation 
history of these systems. 
In the last years, a number of observational studies were produced 
to constrain the origin of PRGs (Reshetnikov \cite{Resh97}; see also the
review by Sparke \& Cox \cite{Cox2000}).
In almost all PRGs the morphology of the host galaxy resembles that of
an early-type object (elliptical or S0 galaxy), because of its structure-less 
appearance and no HI: kinematical studies on
some PRGs have confirmed that this component is rapidly rotating
(Schechter et al. \cite{Schechter84}; Whitmore et al. \cite{PRC}).
The integrated colors and gas-to-dust ratio, together with the
large $M(HI)/L_B$ ratio for the whole system suggest that PRGs may be 
quite similar to the late-type spirals (Arnaboldi et al. \cite{magda95}).
Very recent works on NGC~4650A, which is considered the prototype of the class
of wide PRGs, (Arnaboldi et al. \cite{magda97};
Iodice et al. \cite{4650aI}; Gallagher et al. \cite{4650aG}) 
have shown that the polar structure appears to be a disk of a very young
age; moreover the host galaxy integrated colors and light
distribution do not resemble that of a typical early-type system. 

The main goal of the present work (Paper I) is to provide accurate photometry
in the NIR for a sample of PRGs; in a second paper (Iodice et al. \cite{paperII}, 
Paper II) they will be compared with the predictions 
from different formation scenarios for these peculiar systems.
Near-IR photometry is necessary to reduce as much as possible the
dust absorption that strongly affects the starlight distribution in the host
galaxy and in the ring (Whitmore et al. \cite{PRC}). Since the dust optical 
depth decreases toward longer wavelengths, photometry in the NIR
will be relatively free from this problem, and the inner structures
of the host galaxy and ring may be easily identified. 
The NIR photometry is also more representative
of the light emitted by the older stellar population, which contains most
of the mass: any dynamical modeling of PRGs
will be more accurate when using the light distribution for the
different components in the NIR rather than those at the optical
wavelength.                               
In addition, the study of optical and NIR integrated colors will yield
information about the age and metallicity of dominant stellar population
in the different components of a PRG system.

In this paper we present new NIR observations, obtained for a 
selected sample of PRGs from the Polar Ring Catalogue, listed in 
Table~\ref{prg}, and have applied the same procedures adopted
to study the polar ring galaxy NGC~4650A (Iodice et al. \cite{4650aI}).
Observations and data reduction are presented in Sec.\ref{observations}; 
the morphology, light and color distribution of the two components 
(host galaxy and ring) are discussed in Sec.\ref{morphology} and 
Sec.\ref{photometry}. In Sec.\ref{color} the integrated colors 
derived for different regions of each PRG are presented. 
The two-dimensional model of the host galaxy light distribution is discussed in 
Sec.\ref{2Dmodel},a detailed description of each selected
PRG is given in Sec.\ref{obj_descr}. The final summary of the data is 
presented in Sec.\ref{conclu}.                                               

\begin{table*}
\centering
\caption[]{The Polar Ring Galaxy sample studied in this work. 
In the second column of the table we list the  object identification as
given in the Polar Ring Catalogue, PRC, (Whitmore et al. \cite{PRC}); 
coordinates $\alpha$ and $\delta$, the Heliocentric velocities, 
and galaxy extension (derived from NED database) are reported in the third, fourth, 
fifth and sixth columns respectively.}
\label{prg}
\begin{tabular}{cccccc}
\hline\hline
Object name & PRC name & $\alpha$ (J2000) & $\delta$ (J2000) & $V_{0}$ & diameters\\
  &  &  &  & (km/s) & (arcmin)\\
\hline
A0136-0801 & A-01 & 01h38m55.2s & -07d45m56s & 5500 & 0.41 x 0.3\\
ESO 415-G26 & A-02 & 02h28m20.1s & -31d52m51s & 4604 & 1.3 x 0.6 \\
ARP 230 & B-01 & 00h46m24.2s & -13d26m32s & 1742 & 1.3 x 1.2\\
AM 2020-504 & B-19 & 20h23m54.8s & -50d39m05s & 4963 &   \\
ESO 603-G21 & B-21 & 22h51m22.0s & -20d14m51s & 3124 & 1.1 x 0.6 \\
\hline
\end{tabular}
\end{table*}      

\section{Observations and data reduction}\label{observations}
The near-infrared J, H  and Kn data were acquired during several 
observing runs at the 2.3 m telescope of the Mt. Stromlo and Siding Spring 
Observatory, with the CASPIR infrared camera (McGregor \cite{McG94}), by 
M. Arnaboldi and K.C. Freeman. 
The angular resolution of the data is 0.5 arcsec pixel$^{-1}$ and covering
a field of view of $2.0' \times 2.0'$.
The observing log for these data is summarized in Table~\ref{obslog}.
Images were acquired in the offsetting mode: a cycle was defined containing
5 images on target, interspersed with 5 sky frames; each object frame was taken 
with a small offset from the galaxy center and the sky frames were taken before 
and after each galaxy frame. Dark frames were acquired at the beginning and end
of each cycle, while bias frames were obtained at the beginning and end of 
each set of cycles. 
More cycles were obtained in the Kn band than in the J and H band, in order
to have a better estimate of the background level.
For A0136-0801 only the H band observations are available.
The data reduction is carried out by using the CASPIR package in the 
IRAF\footnote{IRAF is distributed by the National Optical Astronomy 
Observatories,  which is operated by the Associated Universities for
Research in Astronomy, 
Inc. under cooperative agreement with the National Science Foundation.} 
({\it Image Reduction and  Analysis Facility}) environment.
The main strategy adopted for each data-set includes linearization,  
flatfielding 
correction, sky subtraction and rejection of bad pixels. Finally,
all frames were registered and co-added into the final scientific frame.
Several standard stars, from Carter \& Meadows (\cite{carter95}), were  
observed at the beginning, middle and at the end of each observing night,
in order to transform
instrumental magnitudes into the standard J, H and Kn band systems.
All the objects of the sample are small: they subtend a diameter of
about $\sim 1'$ on the sky and therefore they lie well inside the detector
field, and allow for a suitable estimate of the sky background. 
The calibrated H-band images of each object are shown in 
Fig.~\ref{hfm1} (left panels).

\begin{figure}[h]
\centering
\includegraphics[width=6cm]{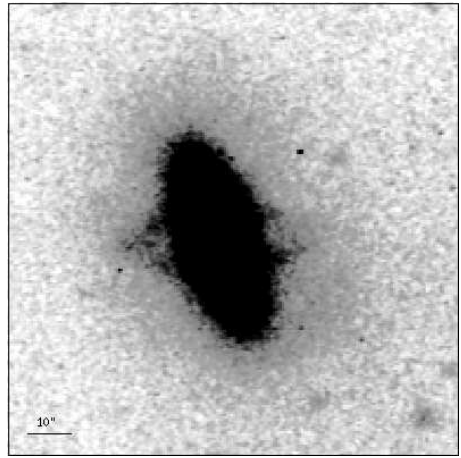}
\caption{B band images for ESO~415-G26. Units are intensity; North is up and East 
is to the left.}  
\label{eso415_B}
\end{figure}

To study the stellar population ages of the two components in PRGs 
(see Paper II), we need the optical data for all the objects of our
sample.
The B band images for the polar ring galaxies ESO~603-G21,
AM~2020-504 and ARP~230, used in the present work were obtained
by Arnaboldi and collaborators (Arnaboldi et al. \cite{magda95}).
For ESO~415-G26, new B band data were acquired in January 2001,
by K.C. Freeman, at the Mt. Stromlo and Siding Spring Observatory (SSO) 2.3 m 
telescope.
The imager at the SSO 2.3 m telescope has a field of view of $10' \times 
10'$ and an angular resolution of 0.59 arcsec pixel$^{-1}$.
The data-set included 3 bias frames, 3 different images of the twilight
sky (to derive the flatfield image) and 5 object frames taken with the
offsetting mode.
The total integration time of the object frames is 1200 seconds.
Moreover, standard stars in the E2-region (Graham, \cite{graham82}) 
wer observed in B and  V bands in order to establish the photometric zero
point for the B band system.
To obtain the final B band image for ESO~415-G26 (Fig.\ref{eso415_B}), 
the data reduction is similar to that described for the NIR images. 
The CCDRED package in IRAF was used
to linearise, flatfield and combine these B band images.

\begin{table}
\centering
\caption[]{NIR log of observations for the selected sample of polar ring 
galaxies.}
\label{obslog}
\begin{tabular}{ccccc}
\hline\hline
Object & Filter & Tot. int. & FWHM & Date \\
       &        &  (s)      &(arcsec)&    \\
\hline
A0136-0801 & H  & 900& 3.7& 20/08/1995\\
ESO 415-G26 & J & 600& 3.1 & 20/08/1995\\
            & H & 600 & 1.9 & 19/08/1995\\
            & H & 600 & 2.3 & 20/08/1995\\
            & Kn & 900 & 1.4 & 18/08/1995\\
            & Kn & 300 & 1.3 & 19/08/1995\\
ARP 230 & J & 900 & 1.5 & 19/08/1995\\
        & H & 900 & 1.6 & 19/08/1995\\
        & Kn & 2100 & 1.6 & 18/08/1995\\
AM 2020-504 & J & 1200 & 2.2 & 18/08/1995\\
            & H & 1200 & 1.4 & 18/08/1995\\
            & Kn & 900 & 1.8 & 19/08/1995\\
            & Kn & 1200 & 2.0 & 20/08/1995\\
ESO 603-G21 & J & 900 & 1.5 & 19/08/1995\\
            & H & 1200 & 1.6 & 19/08/1995\\
            & Kn & 1800& 1.5 &20/08/1995\\
\hline
\end{tabular}
\end{table}

\section{Host galaxy and polar ring morphology}\label{morphology}
\begin{figure*}
\centering
\includegraphics[width=15cm]{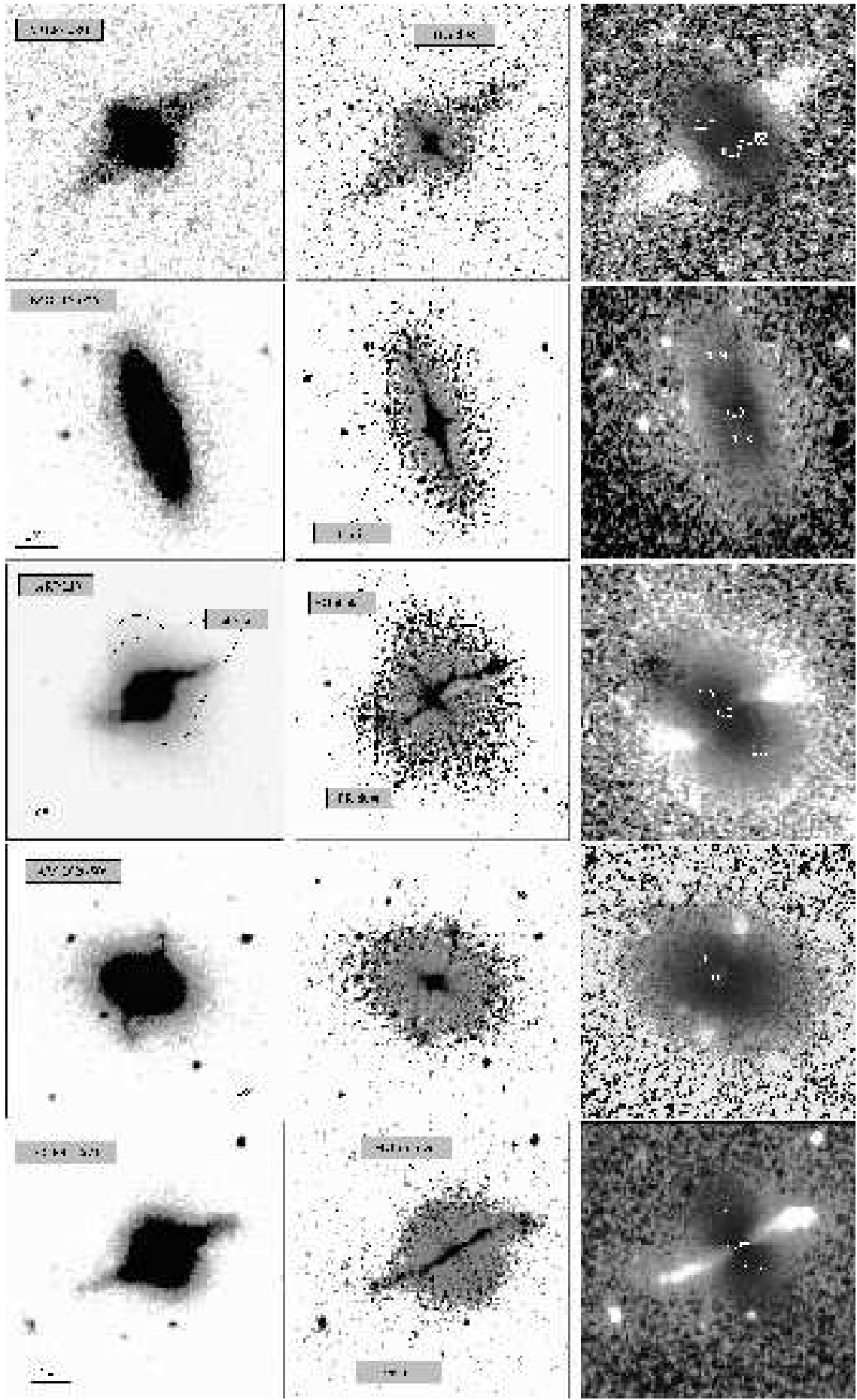}
\caption{H band images (left panels) and relative high-frequency residual  
images (middle panels) for all PRGs of our sample.
Darker colors correspond to brighter features. The image ratio between the
whole galaxy image and the host galaxy model in the H band are
shown in the right panels.
Lighter points correspond to those regions where the galaxy is brighter 
than the model.
Numbers on the right panels indicate the value of the ratio at the 
corresponding positions. 
In all images, North is up and East is to the left.}
\label{hfm1}
\end{figure*}

The NIR images of all objects in the sample show that most of the NIR  
light comes from the host galaxy and its morphology resembles that of a
flattened ellipsoid, most likely a lenticular S0 galaxy (see left panels
of Fig.\ref{hfm1}). In all the objects, but A0136-0801, the polar ring 
is within the optical radius of the central galaxy. 
In A0136-0801 (Fig.\ref{hfm1}, left panel),
the polar ring is more extended in radius than the host galaxy, it has a 
considerable width and hosts several star formation regions, as suggested 
by its irregular light distribution.

We wish to identify the high frequency residuals with respect to the 
homogeneous light distribution. To this aim, we produce a median filtered
image (computed with the 
FMEDIAN package in IRAF) where each 
original pixel value is replaced with the median value in a sliding rectangular 
window. After several tests, a window size of $7 \times 7$ pixels was chosen 
because it provides the optimum enhancement of the S0 inner structures. 
The final un-sharp masked images are obtained as ratios between the
coadded galaxy frame and its median filtered image: this ratio represents
the ``high-frequency residual image'' and it is produced for all NIR
bands (see Fig.\ref{hfm1}, middle panels, for the H band).

The most important result obtained from this analysis is the
identification of a
disk-like structure along the major axis of the host galaxy
for all objects, but AM~2020-504 (Fig.\ref{hfm1}). 
Furthermore the high-frequency residual images in ESO~603-G21 and ARP~230 reveal 
additional fainter structures in the inner regions of the host galaxy,
which are probably related to the polar ring (Fig.\ref{hfm1}).
All these luminous components become stronger in the H and Kn bands,
as the perturbations due to dust absorption decrease in these bands.

The resolution of our groundbased NIR data does not allow us 
to establish whether the polar structure extends all the way in to the
galaxy center, as Iodice et al. (\cite{4650aI}) found in NGC~4650A.
A0136-0801 is the only wide polar ring in our sample with
a similar morphology to NGC~4560A, but HST data are not available for this galaxy
to perform a similar study as for NGC~4650A.

In Sec.\ref{obj_descr}, the high-frequency residual images for all objects and
in all bands are discussed in details.

\section{Photometry: light and color distribution}\label{photometry}
\begin{figure*}
\centering
\includegraphics[width=13cm]{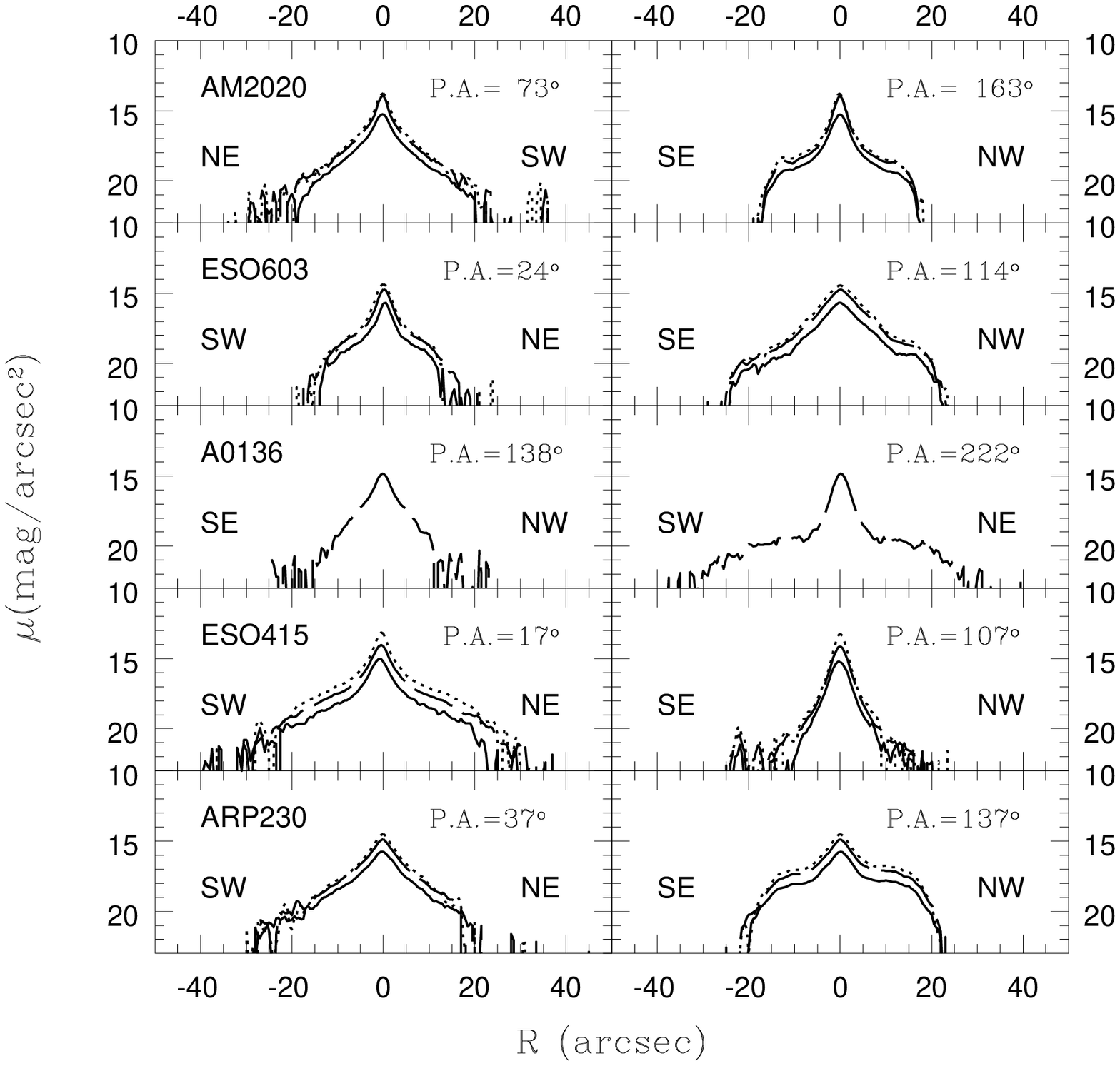}
\caption{J (continuous line), H (dashed line) and Kn (dotted line) light profiles 
along the host galaxy major axis (left panels) and along polar ring major axis 
(right panels) for all PRGs of our sample.} 
\label{prof}
\end{figure*}

For each object, the J, H and Kn luminosity profiles were derived along
the major axis of the host galaxy and are shown in the left panels of 
Fig.\ref{prof}. 
To map the whole polar ring extension, we computed the surface brightness 
profiles for this component as an average of several profiles extracted 
in a cone, centered on the host galaxy and several degrees wide, at the 
Position Angle (P.A.) of the ring major axis (right panels of Fig.\ref{prof}).
The width of the cone depends on the inclination respect to the line-of-sight
and the radial extension of the ring.  

\begin{figure*}
\centering
\includegraphics[width=15cm]{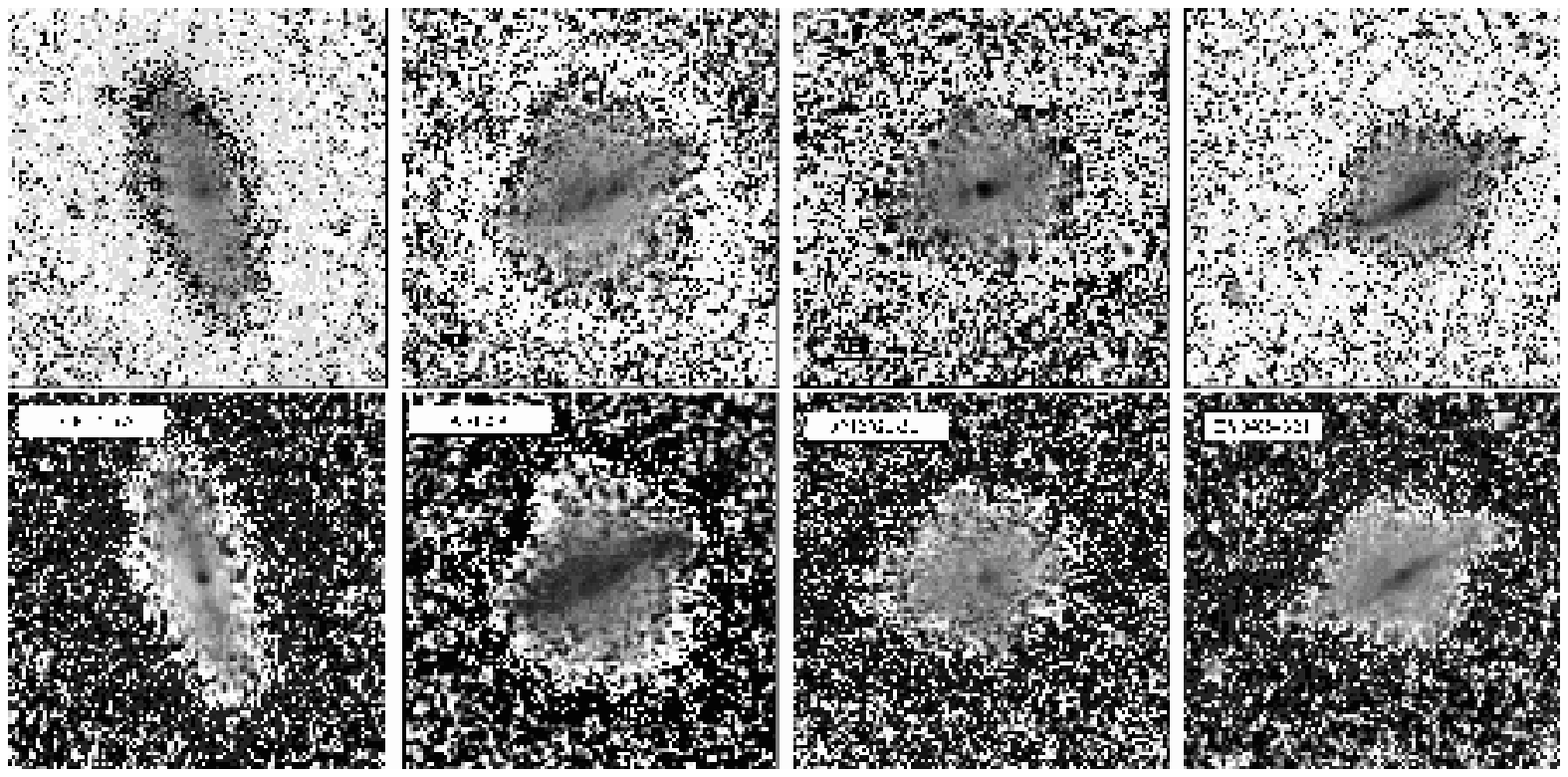}
\caption{J-H (top panels) and H-K (bottom panels) color maps
for all PRGs of our sample.
Color code in each map is: darker regions correspond to redder colors, and
lighter regions correspond to bluer colors.
North is up and East is to the left.}
\label{mappe}
\end{figure*}

The J-H and H-K color maps, derived for ESO~415-G26, ARP~230, AM~2020-504 and 
ESO~603-G21, are shown in Fig.\ref{mappe}:
in all PRGs, but AM~2020-504, there is a redder region which corresponds
to the disk-like component already identified in the high-frequency images 
(Sec.\ref{morphology}). 
The characteristics of the light and color distributions in each object will be
discussed in detail in Sec.\ref{obj_descr}.

\section{Integrated magnitudes and colors}\label{color}

For each PRG of the sample, the total integrated magnitudes are computed 
in the NIR, J, H and Kn bands, inside a circular aperture of a given radius. 
Such radius depends on the average extension of the object, which can be 
easily deduced from the light profiles shown in Fig.~\ref{prof}.
Total magnitudes were corrected for the extinction within
the Milky Way, by using the absorption coefficient in the B band 
($A_B$) and the color excess $E(B-V)$ derived from Schlegel et
al. (\cite{Schlegel98}).
The absorption coefficients for the J, H and Kn bands ($A_J$, $A_H$, $A_{Kn}$) 
are derived by adopting $R_V=A_V/E(B-V)=3.1$, and using the $A_\lambda$
curve from Cardelli et al. (1989). 
The values of the absorption coefficients $A_\lambda$ adopted for each object 
are listed in Table~\ref{abs}.
Total magnitudes are listed in Tab.~\ref{totm}; furthermore for A0136-0801, 
ARP~230 and ESO~603-G21 the table includes the total magnitudes derived from the 
Two Micron All Sky Survey (2MASS) data, in the same regions.
The average uncertainties which affects the 2MASS magnitudes listed in 
Tab.~\ref{totm} varies between 0.02 and 0.04 mag. However, these values
are computed for the Poissonian noise, and they do not include 
additional photometric uncertainties\footnote{Atmospheric OH airglow emission
variations contribute extra noise particularly in the H band, where
the total photometric error can be about 0.15 mag.} 
which affect these data and cause the total photometric error to be about three
times larger (Jarrett et al. \cite{2mass}). 

The J, H and Ks total magnitudes derived from the 2MASS data
are on average 0.16 mag brighter than the J, H and Kn CASPIR 
magnitudes (see Tab.\ref{totm}). 
The non-Poissonian background fluctuations, which affects both CASPIR
and 2MASS images, particularly in H band, can only account for such differences.

\begin{figure*}
\resizebox{\hsize}{!}{
\includegraphics[width=5cm]{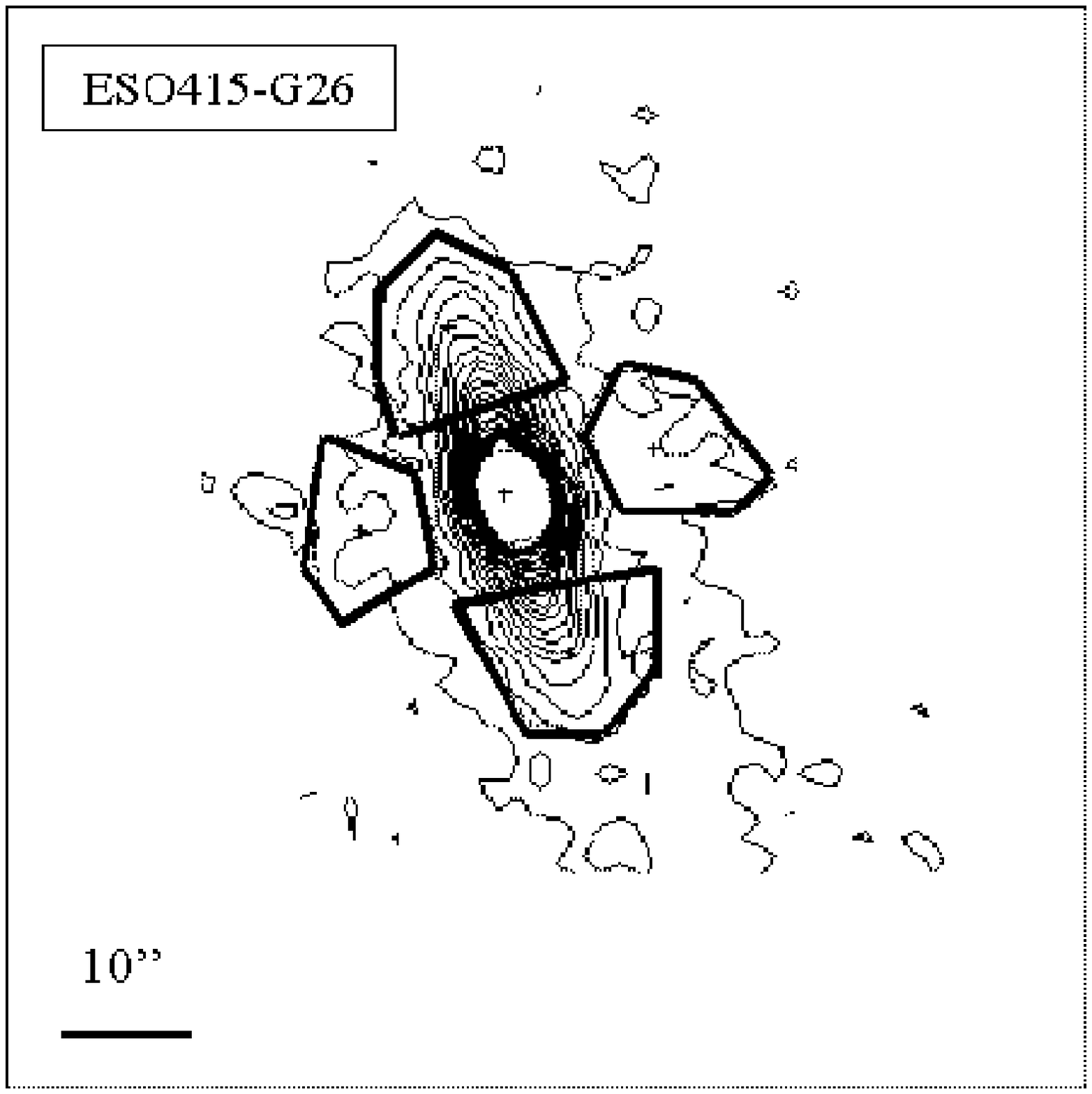}
\includegraphics[width=5cm]{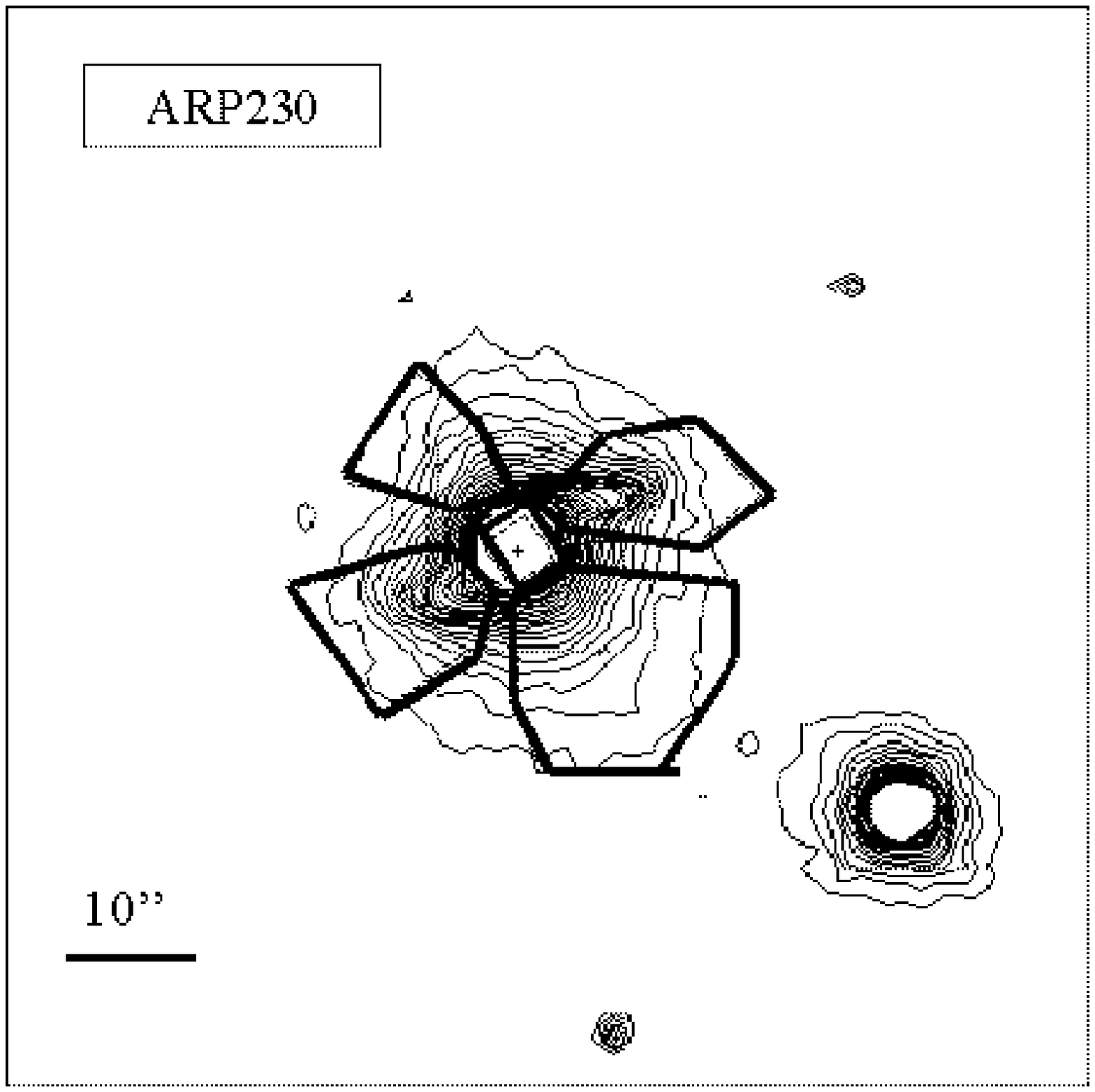}}
\resizebox{\hsize}{!}{
\includegraphics[width=5cm]{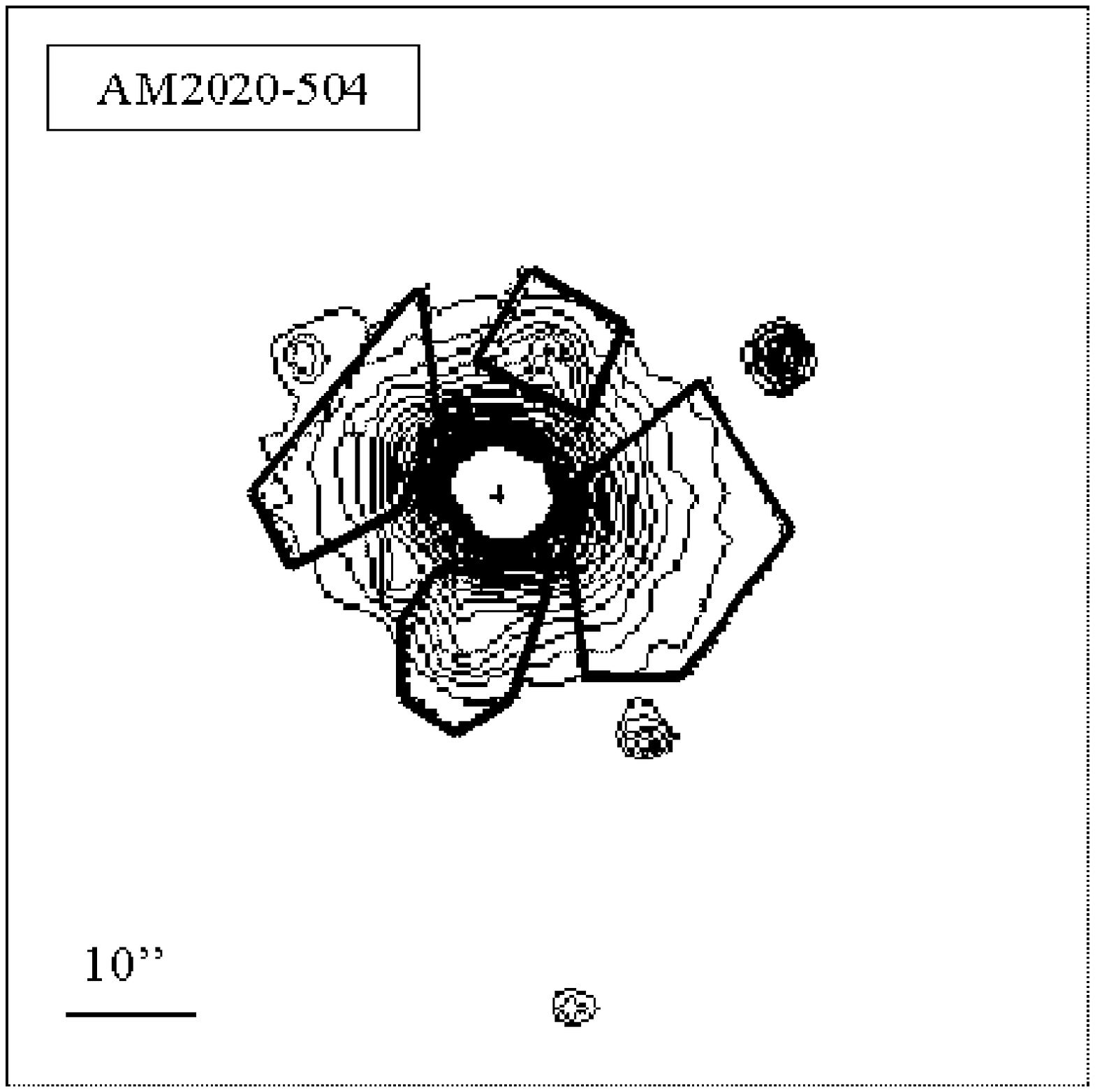}
\includegraphics[width=5cm]{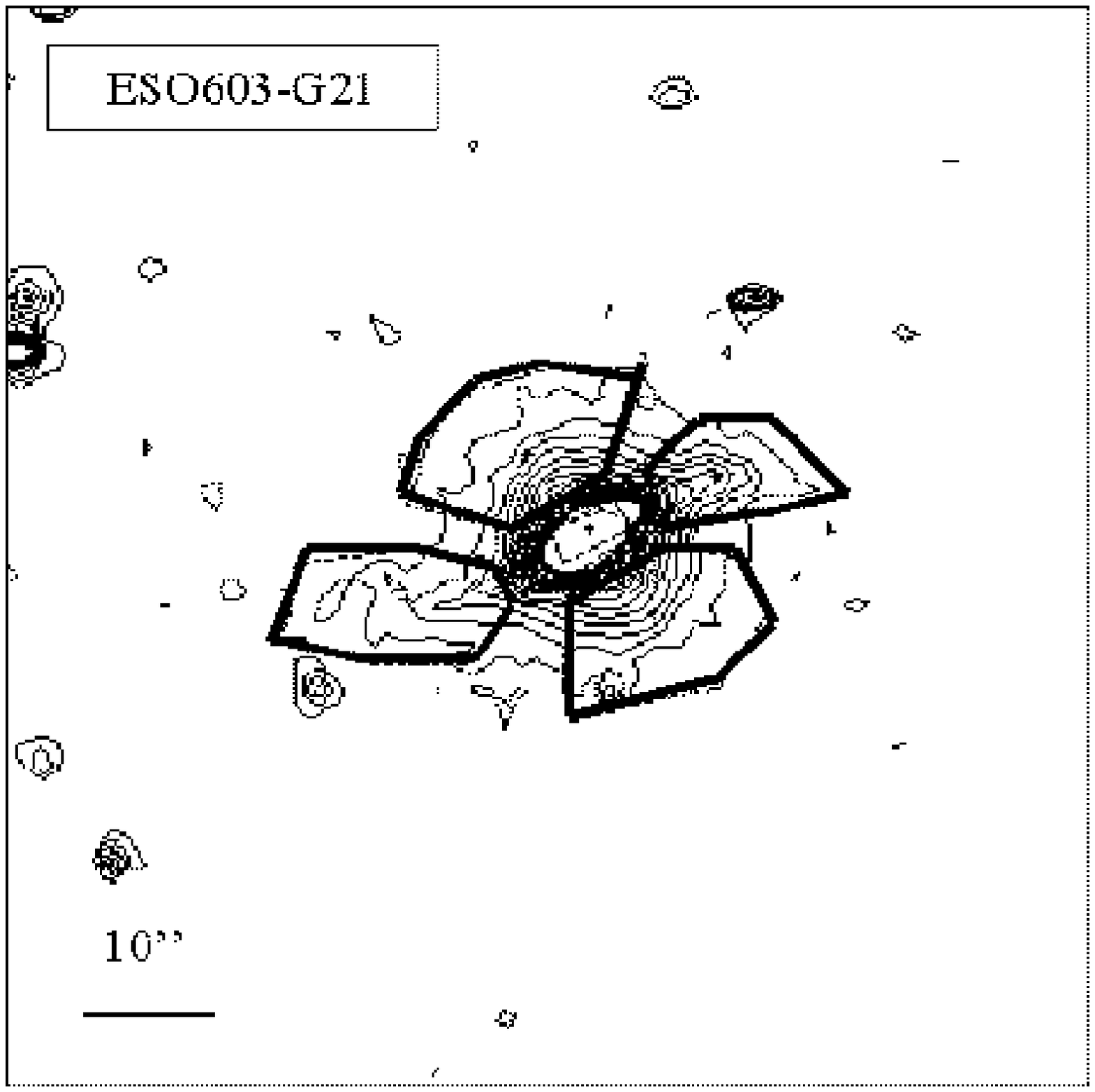}}
\caption{Intensity contour plot in the J band plus the five polygons
limiting the different areas where the integrated magnitudes are computed 
(heavier lines) for ESO~415-G26 (top right panels), ARP~230 (top left panels), 
AM~2020-504 (bottom right panels) and ESO~603-G21 (bottom left panels).
North is up and East is to the left.} 
\label{poly}
\end{figure*}

In order to compute a detailed analysis of the stellar population content of 
the main components in PRGs (host galaxy and polar ring) we have derived the 
integrated magnitudes and colors in five different areas.
The five areas are chosen as follows: one is coincident with the nucleus;
two areas are placed within the host galaxy stellar component, but outside
the nucleus and in regions unperturbed by the polar ring), and two areas 
for the polar ring. 
In Fig.\ref{poly} these regions are shown for each object of the sample.
The polygon contours limit these areas and are determined in the J
image; the same polygons are used for H and Kn bands, after the images
where registered to the J image. 
Polygons are marked by using the IRAF task POLYMARK and the integrated 
magnitude are computed using the IRAF task POLYPHOT. 
A detailed discussion on the photometric errors is given in 
Appendix\ref{error}.
The integrated magnitudes and colors derived in each area
were corrected for the extinction within
the Milky Way, by using the absorption coefficients listed in Table~\ref{abs}.
The corrected integrated magnitudes and colors are listed in Table~\ref{polycol}.

\begin{table}[h]
\centering
\caption[]{Absorption coefficients for the Galactic extinction in the 
direction of PRGs.}
\label{abs}
\begin{tabular}{lcccc}
\hline\hline
Object & $A_B$ & $A_J$ & $A_H$ & $A_{Kn}$ \\
       & (mag) & (mag) & (mag) & (mag)\\
\hline
A0136-0801 & 0.115  & 0.024 & 0.015 & 0.01\\
ESO 415-G26 & 0.073 & 0.015 & 0.01 & 0.006\\
ARP 230 & 0.083 & 0.017 & 0.011 & 0.007\\
AM 2020-504 & 0.185 & 0.039 & 0.025 & 0.016\\
ESO 603-G21 & 0.144 & 0.030 & 0.019 & 0.012\\
\hline
\end{tabular}
\end{table}

\begin{table*}[h]
\centering
\caption[]{Total magnitudes inside a circular area of a given radius {\it r}.
The total magnitudes $m_J^{2M}$, $m_H^{2M}$ and $m_{Ks}^{2M}$ are derived 
from the 2MASS images.}
\label{totm}
\begin{tabular}{lccccccc}
\hline\hline
Object& {\it r}&$m_J \pm 0.04$ &$m_J^{2M}$& $m_H \pm 0.06$& 
$m_H^{2M}$& $m_{Kn} \pm 0.04$& $m_{Ks}^{2M}$ \\
      & (arcsec)& (mag)& (mag)& (mag)& (mag)& (mag)& (mag)\\
\hline
A0136-0801 & 40&      & 12.90& 12.62& 12.49&      & 12.05\\
ESO 415-G26& 30& 12.48&      & 11.86&      & 11.38&      \\
ARP 230    & 25& 12.00& 12.00& 11.41& 11.31& 11.14& 10.96\\
AM 2020-504& 25& 12.50&      & 11.83&      & 11.60&      \\
ESO 603-G21& 30& 12.79& 12.64& 12.08& 11.88& 11.65& 11.52\\
\hline
\end{tabular}
\end{table*}

\begin{table*}[h]
\centering
\caption[]{Integrated magnitudes and colors of different regions of host 
galaxy and polar ring.}
\label{polycol}
\begin{tabular}{lcccccccc}
\hline
Object&component&region&$m_B$ (mag)&$m_J$ (mag)&B-H&J-H& H-K&J-K\\
\hline\hline
ESO 415-G26 & PR & SE & 17.97 & 16.25& 2.17 &0.45 & 0.36& 0.81\\
            & PR & NW & 17.22 & 15.90& 2.13 &0.32 & 0.41& 0.73\\
            & HG & SW & 17.16 & 14.96& 2.84 &0.64 & 0.45& 1.09\\
            & HG & NE & 16.78 & 14.87& 2.64 &0.72 & 0.51& 1.23\\
            & HG & center & 15.80 & 13.42& 3.32 &0.80 & 0.54& 1.34\\
ARP 230     & PR & SE & 15.96& 14.24& 2.29 &0.57 & 0.29 & 0.85\\
            & PR & NW & 16.24& 14.81& 1.94 &0.51 & 0.29 & 0.81\\
            & HG & SW & 15.71& 13.89& 2.3  &0.48 & 0.20 & 0.67\\
            & HG & NE & 16.25& 15.02& 1.81 &0.57 & 0.19 & 0.76\\
            & HG & center & 16.42& 13.97& 3.21& 0.75 & 0.32& 1.08\\
AM 2020-504 & PR & NW & 18.10& 15.96& 2.54 & 0.57 & 0.27& 0.86\\
            & PR & SE & 17.96& 16.07& 2.32 & 0.60 & 0.19& 0.81\\
            & HG & SW & 18.03& 15.03& 3.40 & 0.60 & 0.17& 0.79\\
            & HG & NE & 18.19& 15.43& 3.22 & 0.63 & 0.21& 0.85\\
            & HG & center & 16.63& 13.50& 3.73 & 0.77 & 0.25& 1.04\\
ESO 603-G21 & PR & SE & 17.71& 15.92& 1.90 & 0.46 & 0.45& 0.91\\
            & PR & NW & 17.68& 14.95& 2.07 & 0.58 & 0.37& 0.96\\
            & HG & SW & 17.29& 14.83& 3.03 & 0.57 & 0.37& 0.94\\
            & HG & NE & 17.24& 14.83& 3.00 & 0.59 & 0.33& 0.92\\
            & HG & center & 18.96& 14.19& 5.71 & 0.94 & 0.40& 1.34\\
\hline
\end{tabular}
\end{table*}

\section{Two dimensional model of the host galaxy light 
distribution}\label{2Dmodel}
One of the open issues in the study of PRGs is the nature
of the host galaxy. A qualitative morphological inspection suggests that
this component is similar to an early-type galaxy, most likely an S0
(Sec.\ref{morphology}). 
In this paper we present the structural parameters from the two-dimensional (2D)
bulge-to-disk decomposition, while the analysis of the results and the
comparison with standard S0 systems will be presented in Iodice et al. \cite{paperII} 
(Paper II). 
                                                    
The 2D model fit to the host galaxy light distribution 
is based on the super-position of two components,
each one characterized by concentric and co-axial elliptical isophotes with
constant flattening (see Iodice et al. \cite{Iodice2001}). 
One of this component may be thought as the projection of a
spheroid with finite intrinsic thickness, i.e. the bulge. The other
component is a disk.           
The projected surface brightness distribution of the spheroidal component 
follows the Sersic law:
\begin{equation}\label{sersic}
\mu_{b}(x,y)=\mu_{e}+k \left[\left(\frac{r_{b}}{r_{e}}\right)^{1/n}-1
\right]
\end{equation}
with $k=2.17n-0.355$ and $r_{b}=\left[{x^{2}+y^{2}/q_b^{2}}\right]^{1/2}$;
$q_{b}$, $\mu_{e}$ and $r_{e}$ are the {\it apparent axial ratio}, the
{\it effective surface brightness} and the {\it effective radius}
respectively. 
The total luminosity of this component is then $L_{B}=K(n) I_{e} r_{e}^2$ 
(Caon et al. \cite{caon93}), where $K(n)$ is a 
function of the $n$ parameter and $I_{e}= 10^{-0.4 \mu_{e}}$.
The Sersic law is able to fit an exponential behaviour when the n 
parameter is close to 1. 

The projected surface brightness distribution of the disk 
follows an exponential behavior, as described by Freeman's exponential 
law (\cite{freeman70}):

\begin{equation}\label{exp}
\mu_{d}(x,y)=\mu_{0}+1.086\left(\frac{r_{d}}{r_{h}}\right)
\end{equation}
with $r_{d}=\left[{x^{2}+y^{2}/q_d^{2}}\right]^{1/2}$; $q_{d}$,
$\mu_{0}$, and $r_{h}$ are the {\it apparent axial ratio},
{\it the central surface brightness} and the {\it scalelength} of the 
disk, respectively. The central surface brightness corrected for the 
inclination of the disk with respect to the line of sight is given by 
$\mu_{0}^{c}= \mu_{0}-2.5\log(q_{d})$, and the total luminosity for this 
component is  $L_{D}=2\pi I_{0} r_{h}^2$, with $I_{0}=10^{-0.4 \mu_{0}^{c}}$ 
(Freeman, \cite{freeman70}). 
The total {\it Bulge to Disk ratio (B/D)} is given by the ratio between the 
total luminosities of the bulge and disk components ($B/D=L_{B} / L_{D}$).

For all PRGs of the sample, the 2D model of the host galaxy light distribution
 is performed in the Kn band, because the dust absorption is weaker in this
band. For the polar ring galaxy A0136-0801 only the H band data were available,
so the study of the light distribution for this object is done in this 
band. However, we have derived the effective surface brightness of the 
bulge and the central surface brightness of the disk in the Kn band by
using the average H-K colors computed from the 2MASS magnitudes, available 
for this system (see Tab.~\ref{totm}). 
Since the host galaxy morphology in AM~2020-504 seems to be more similar to 
that of an elliptical rather than an S0 galaxy (see Sec.\ref{morphology}), 
its light distribution was modeled with a single component, given by the 
Sersic law.  
The regions affected by foreground stars and by the polar ring light
were accurately masked before performing the fit to the light distribution.
Particular attention was given to the seeing effects: to avoid biased
results  caused by seeing, we have masked out the central regions of the
galaxy, before 
performing the fit. The dimension of the masked area is as large
as the measured PSF in that image.
Each galaxy point in the fitting algorithm is weighted by its luminosity,
as brighter points are affected by smaller errors. 
Then the models for the host galaxy in the J and H bands are
the scaled versions of the Kn band model, based on the average colors of 
this component (see Sec.\ref{photometry}).
The structural parameters derived for each object, and the relative 
error estimated by the algorithm are listed in Table~\ref{2dparam}. 
In the last line of Table~\ref{2dparam} we reported the $\tilde{\chi} ^2$ 
for each fit: values for $\tilde{\chi} ^2 \le 2$ represent good fits 
(as described by Schombert \& Bothun \cite{schombert87}).

\begin{figure}[h]
\centering
\includegraphics[width=7cm]{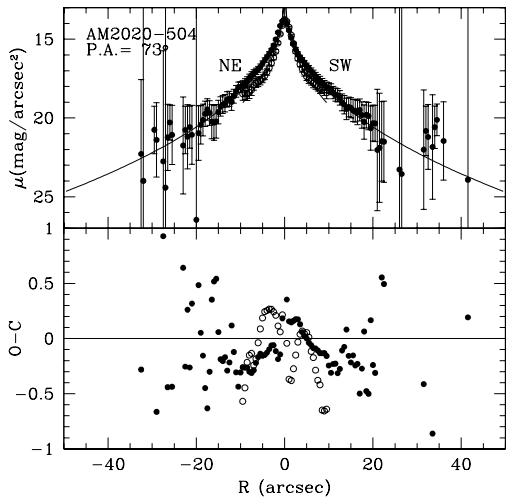}
\caption{2D fit of the host galaxy light distribution for the polar ring
galaxy AM 2020-504; a single component model was adopted.
The observed light profiles along the major (filled dots) and minor axis 
(open dots), in the Kn band, are compared with those derived from the fit.
The orientation and P.A., reported on this figure, refer to the major
axis.}
\label{fit_prof2}
\end{figure}

\begin{figure*}[h]
\resizebox{\hsize}{!}{
\includegraphics[width=6cm]{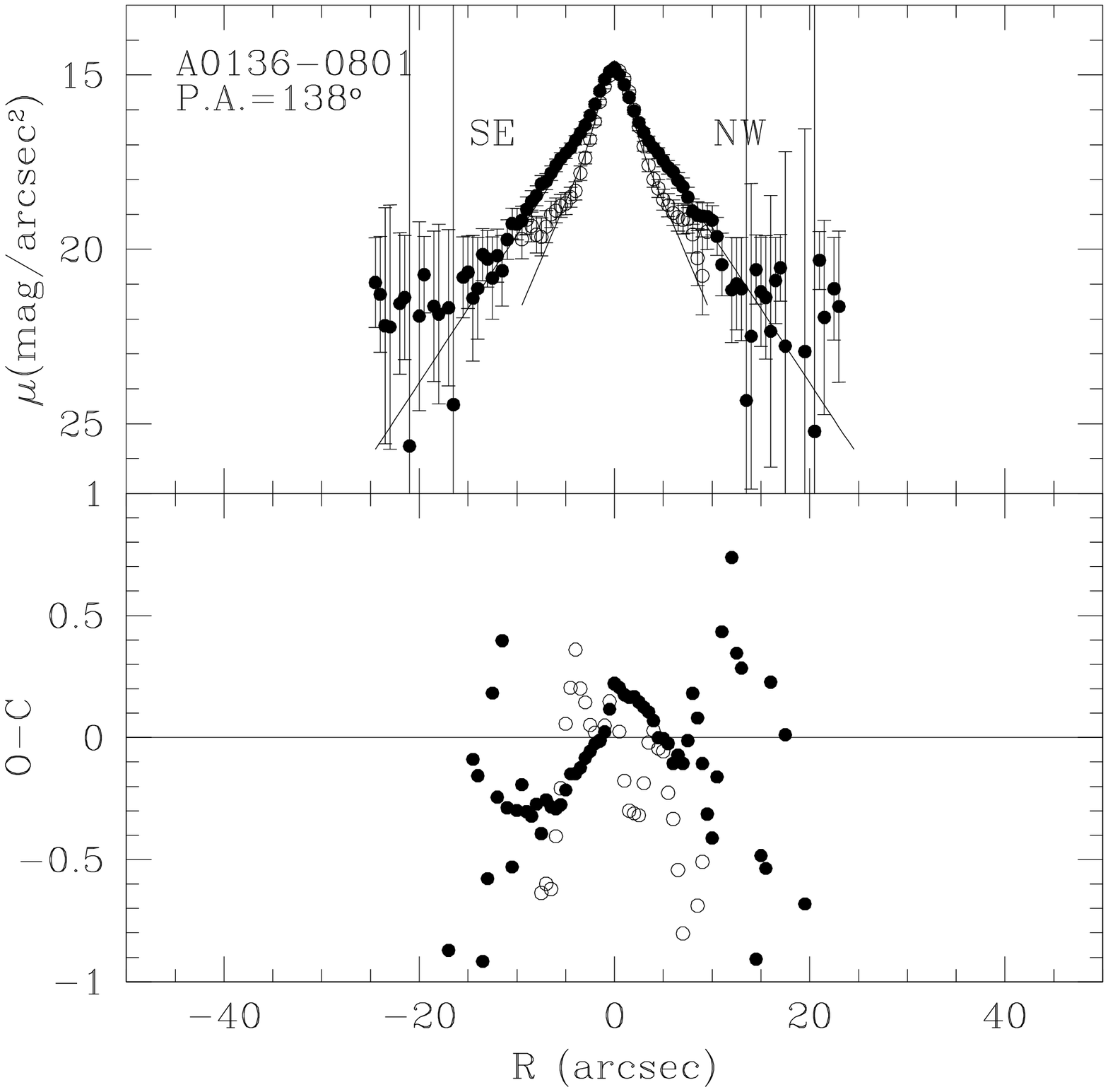}
\includegraphics[width=6cm]{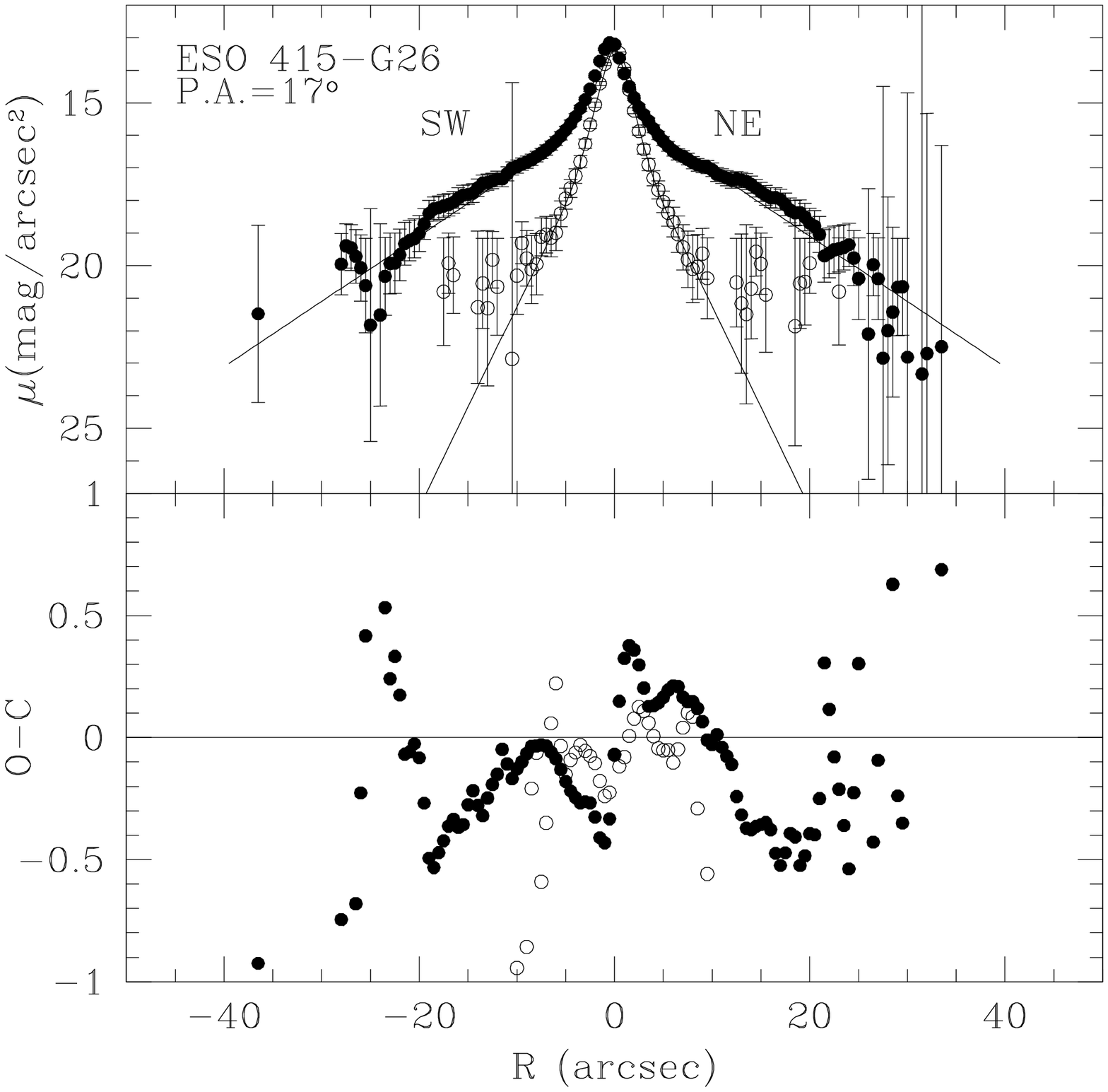}}
\resizebox{\hsize}{!}{
\includegraphics[width=6cm]{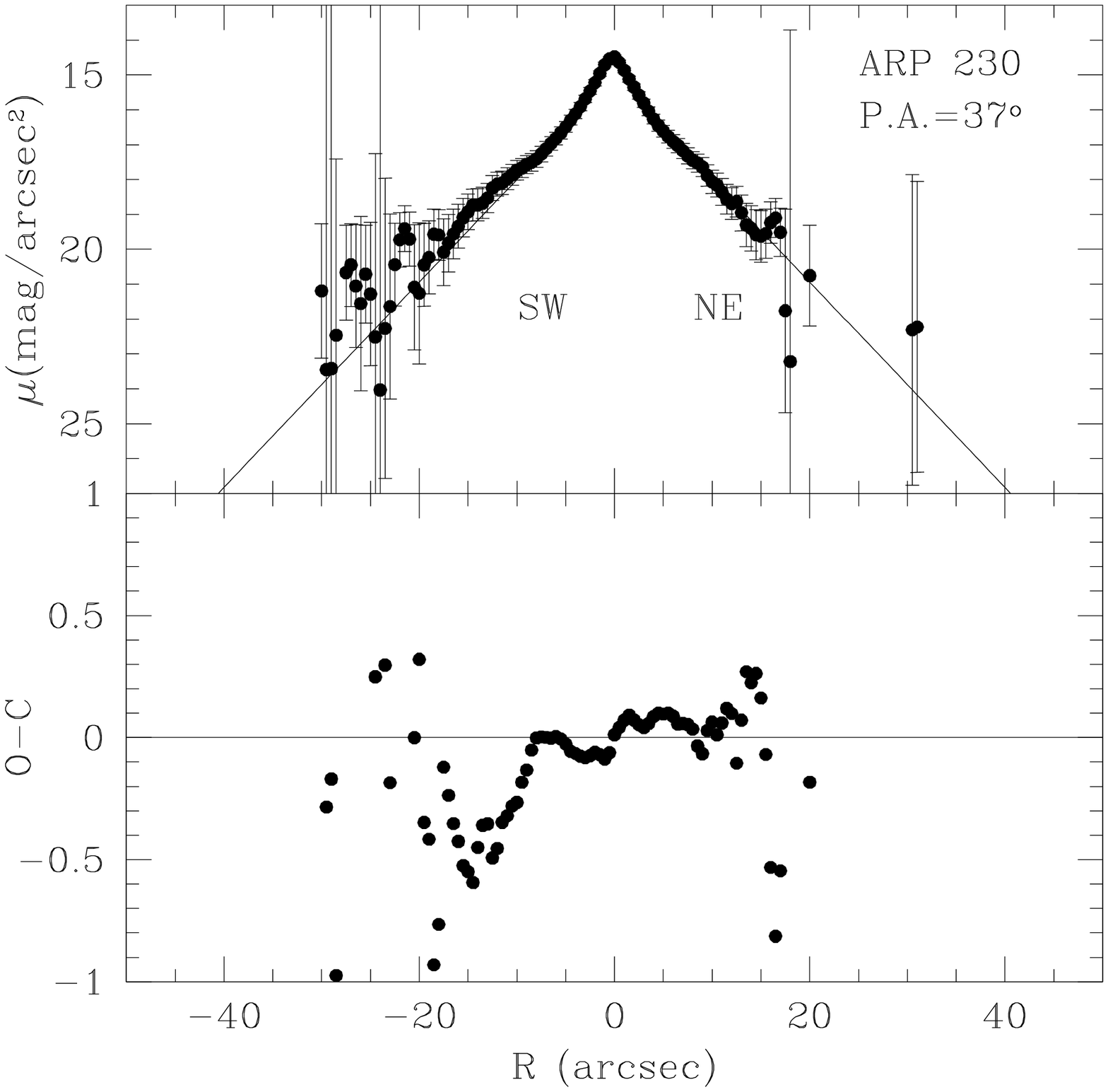}
\includegraphics[width=6cm]{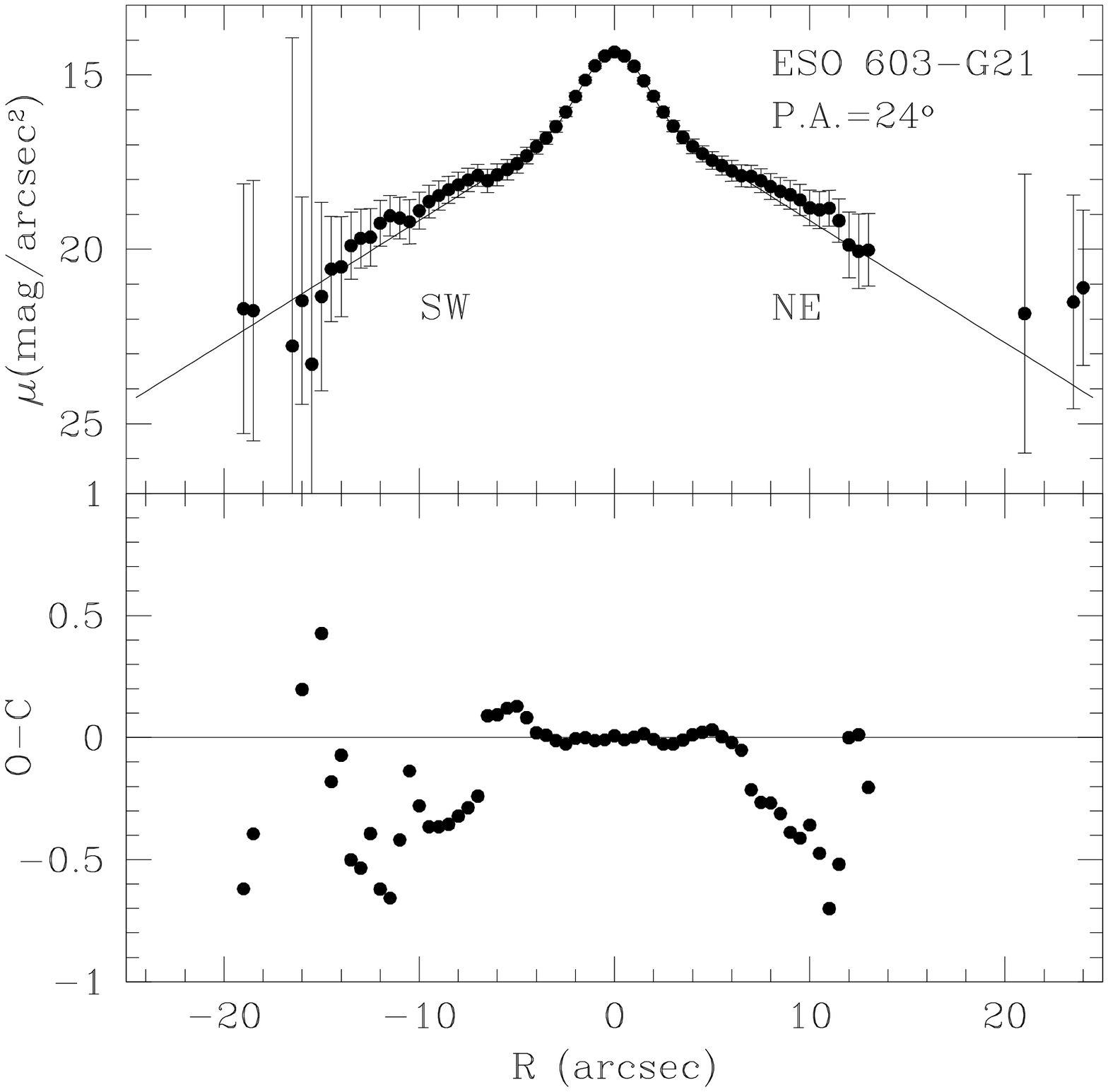}}
\caption{2D fit of the host galaxy light distribution with
the two-component model (bulge+disk), described in Sec.\ref{2Dmodel}.
The observed light profiles along the major (filled dots) and minor axis 
(open dots) are compared with those derived by the fit (continuous line),
which are performed in the Kn bands for ESO~415-G26,
ARP 230, ESO~603-G21 and in the H band for A0136-0801.
For ARP 230 and ESO~603-G21 the luminosity profile is that along the 
orthogonal direction to the polar ring.
The orientation and P.A., reported on each panels, refer to the major axis.} 
\label{fit_prof1}
\end{figure*}

In ESO~603-G21 and ARP 230 the ring radius is not large enough for the PR
to be clearly separated from the host galaxy, (see Fig.\ref{hfm1}): 
here the structural parameters, in particular the apparent 
axial ratios, for these two objects are affected by larger 
errors than for the other PRGs in the sample. 
Fig.~\ref{fit_prof2} and Fig.~\ref{fit_prof1} show the comparison between the 
observed surface brightness profiles and those derived by the fit, 
along the principal axis of the host galaxy, in each PRGs. 
In regions where the central host galaxy light dominates ($R \le 20$ arcsec), 
the modeled surface brightness distribution differs from
the observed one of about $0.2$ magnitude, and $0.5$ 
magnitude in the outer regions.

In Fig.~\ref{hfm1} (left panels) we show the ratio between the 
whole galaxy image and the host galaxy model, which is computed in the 
H band, for each PRG. This ratio is showed in the 
H band in order to allow for an easy comparison of the residuals
from the 2D fit with the higher frequency
residuals from the unsharp-masking of NIR images (Sec.\ref{morphology}). 
In the regions perturbed by dust absorption, due to the ring component, 
the model is brighter than the galaxy, and this ratio has values less than $1$. 
In these figures the polar 
ring structure stands out clearly and some peculiar luminous features 
related to the galaxy, as the outer shells in
ARP 230 (Fig.~\ref{hfm1} bottom panels).

\begin{table*}[h]
\centering
\caption[]{Structural parameters for the host galaxy in five PRGs of the
sample, measured in the Kn band. 
The effective surface brightness $\mu_{e}$ and the central
surface brightness $\mu_{0}$ are in $mag$ $arcsec^{-2}$; $\mu_{0}^{c}$
is the central surface brightness corrected for the inclination; 
$r_{e}$ and $r_{h}$ are respectively
the effective radius and disk scalelength derived  in arcsec, the corresponding
values expressed in kpc are derived by using $H_{0}=70$km s$^{-1}$ Mpc$^{-1}$.}
\label{2dparam}
\begin{tabular}{cccccc}
\hline\hline
Parameter & A0136-0801 & ESO 415-G26 & ARP 230 & AM 2020-504 & ESO 603-G21\\
\hline
$\mu_{e}$ &$15.83\pm 0.04$& $14.72\pm 0.11$& $16.3\pm 0.2$& $16.61\pm 0.08$& $15.50\pm 0.09$\\
$r_{e}$ (arcsec) &$1.27\pm 0.02$& $1.68\pm 0.07$& $1.7\pm 0.2$& $4.8\pm 0.2$& $1.35\pm 0.08$\\
$r_{e}$ (kpc) &$0.49\pm 0.01$& $0.54\pm 0.02$& $0.20\pm 0.02$& $1.66\pm 0.07$& $0.29\pm 0.02$\\
$q_{b}$ &$0.899\pm 0.005$& $0.80\pm 0.012$ & $0.7\pm 0.5$& $0.75\pm 0.01$& $0.8\pm 0.4$\\
$n$ &$0.59\pm 0.03$& $0.72\pm 0.06$ & $0.6\pm 0.2$& $2.36\pm 0.15$& $0.53\pm 0.14$\\
$\mu_{0}$ &$14.91\pm 0.03$& $15.08\pm 0.02$& $15.06\pm 0.11$& & $15.67\pm 0.15$\\
$\mu_{0}^{c}$ &$15.39\pm 0.04$& $16.30\pm 0.03$ & $15.6\pm 0.9$ & & $15.9\pm 0.9$ \\
$r_{h}$ (arcsec) &$2.56\pm 0.03$& $5.41\pm 0.04$& $3.7\pm 0.2$& & $3.1\pm 0.2$ \\
$r_{h}$ (kpc) &$0.98\pm 0.01$& $1.73\pm 0.01$& $0.45\pm 0.02$& & $0.67\pm 0.04$\\
$q_{d}$ &$0.64\pm 0.007$ & $0.32\pm 0.003$ & $0.6\pm 0.4$& & $0.7\pm 0.5$ \\
$B/D$ &$0.25\pm 0.01$& $0.7\pm 0.4$ & $0.15\pm 0.14$ & &$0.4\pm 0.2$ \\
$\tilde{\chi ^2}$ &1.4& 1.3& 1.5& 1.1& 1.6\\
\hline
\end{tabular}
\end{table*}

\section{Description and discussion of each galaxy}\label{obj_descr}
Here we review in detail the main features which characterize
the light and color distribution in the NIR for each PRG in our sample. 
We will also review other important properties of these objects 
reported in literature, which we will use in Paper II.
\bigskip

{\bf A0136-0801} - 
This is one of the best case of kinematically confirmed polar ring galaxy 
(Schweizer et al. \cite{Schweizer83}; Whitmore et al. \cite{PRC}, PRC): 
it is characterized by a wide polar structure, which is
three times more extended than the optical radius of the central host  
galaxy. 
Along the polar ring major axis the surface brightness profile in the H
band (see Fig.\ref{prof}) is given by two exponential segments 
with different slopes: 
the light inside $10''$ from the center is associated with the host
galaxy, whereas the ring component is extended out to $40''$. Both the H
band image and luminosity profiles (Fig.\ref{hfm1} and
Fig.\ref{prof}) show that
the polar ring is less luminous than the central galaxy, and the bulk of H 
band light is concentrated at smaller radii. The high-frequency residual image 
in the H band (see Fig.\ref{hfm1}) shows the presence of a nearly edge-on 
disk in the central host galaxy.
From the 2MASS database we have derived the J, H and Ks data for this
object and we have computed the total integrated magnitudes in a circular
area, with a $40''$ radius in order to include the whole ring.
This is the same area used to compute the total magnitude in CASPIR H band
(see Tab.~\ref{totm}). Due to the very short exposure (1.3 sec), in the 2MASS
images the polar structure is very faint, and most of the light comes from
the central galaxy. Thus the J-H and H-K colors derived from these data
($J-H=0.41$, $H-Ks=0.43$) are more rapresentative of the host galaxy,
and they are comparable with the J-H and H-K colors obtained for
host galaxy of other PRGs in our sample.
This object was mapped in HI with the {\it Very Large Array (VLA)} by 
van Gorkom et al. (\cite{vgorkom87}) and Cox (\cite{Cox96}): 
all HI emission is found within the polar ring, whose outer HI contours
appear to warp away from the poles.
The total HI mass estimated for this object is about 
$1.6 \times 10^9 M_{\odot}$ (Cox \cite{Cox96}).
The regular HI distribution and optical appearance, an apparent lack of HII 
regions and  other signs of recent star-formation activity (Mould et
al. \cite{Mould82}) suggest that the polar structure is quite old and
possibly dynamically stable.

\bigskip

{\bf ESO 415-G26} -
This is a well-known polar ring galaxy.
Unlike A0136-0801, the polar ring is less extended than the central host 
galaxy in the optical band (see Fig.\ref{eso415_B}). 
Deeper exposures show a lot of debris at a position
angle which is intermediate between those of the host galaxy major axis
and the ring. They also show 
shells and loops in the outer regions (Whitmore et al. \cite{whitmore87}). 
In the NIR bands (Fig.\ref{hfm1}, middle panels) 
the polar ring is so faint that it is hardly detected. 
The central host galaxy is the dominant luminous component:
the analysis of the light distribution shows that 
this is a nearly edge-on S0 galaxy with an exponential bulge. 
Among the PRGs of our sample, ESO~415-G26 is characterized by a highest
B/D ratio (see Table~\ref{2dparam}).
The NIR color maps (Fig.\ref{mappe}) and J-H vs. H-K colors in
different regions of host galaxy and ring  (see Sec.\ref{color})
show that the nucleus of the system is characterized by the reddest color and 
that the polar ring is much bluer than the central host galaxy. The 
HI map for this object was obtained with the VLA by van Gorkom et al. \cite{vgorkom87} 
and by van Gorkom \& Schiminovich \cite{vgorkom97}. 
They noted that the neutral hydrogen lies along the major axis of the polar 
ring, with some degrees of correlation between the HI and the outer
shells (van Gorkom \& Schiminovich \cite{vgorkom97}).
The most accurate estimate of the total HI mass is about 
$5.6 \times 10^9 M_{\odot}$ (Schiminovich et al. \cite{Schiminovich97}).
This object is characterized by a considerable amount of molecular hydrogen:
Galletta et al. (\cite{galletta97}) estimated the total $H_{2}$ mass to
be about $2.4 \times 10^9 M_{\odot}$.

\bigskip

{\bf ARP 230} - 
This object, also known as IC 51, was studied by Wilkinson et
al. (\cite{wilkinson87}) as a well-known example of {\it shell
elliptical} galaxies: in the NE and SW directions, 
outer shells are clearly visible (Hernquist and Quinn \cite{Hernquist88}), 
which are more luminous in the B band 
(see Whitmore et al. \cite{PRC}, PRC) than in the NIR ones
(Fig.\ref{hfm1},  bottom 
panels). It is also classified as a PRG because it has a fast rotating  
disk-like structure, made up by gas, stars and dust, perpendicular to the 
apparent major axis of the central galaxy, $P.A.=37^{\circ}$,  (Mollenhoff
et al. \cite{Mollenhoff92}). Both optical and NIR images (Fig.\ref{hfm1},
bottom panels) of this polar ring  galaxy show that the ring-like
component, along SE and NW directions, has the size of the inner galaxy, 
and it has a very well-defined outer edge, where dust absorption is
present. 
The high-frequency residual images (Fig.\ref{hfm1} right panels), 
show a very distorted  structure for the ring component: 
it seems strongly warped at about $10''$ from the 
center, with associated absorption features.
An elongated structure nearly orthogonal to the ring suggests that the 
central host galaxy is more similar to a disk galaxy, an S0, than an 
elliptical galaxy.
In this case it is very 
difficult to distinguish the morphology of the central component. 
The 2D model of the light distribution also suggests that it may be an S0 
galaxy with an exponential bulge (Table~\ref{2dparam}), but the peculiar ring 
structure produces a very uncertain estimate of all the structural
parameters; the apparent axial  ratios at larger radii are also influenced
by the presence of the outer shells. 
The bright edges of the ring (in the NW and SE directions) and outer 
shells are clearly visible as bright residuals in the image ratio between 
the whole galaxy and the 2D model of the central 
component (see Fig.\ref{hfm1}, lower left panels).
The central host galaxy shows similar J-H colors to the polar ring component, 
and bluer H-K colors (see Table~\ref{polycol}). 
The very red colors of the nucleus of the system are possibly due to dust 
absorption in the ring, which passes in front it.
Near-IR images, in J, H and Ks bands, are also available for this object 
in the 2MASS database. Taking into account the average uncertainties which
affect these data (see Sec.\ref{color}), the 2MASS total magnitudes 
are comparable with those derived from the CASPIR data (see Tab.\ref{totm}.
This object was mapped in HI of the VLA by Schiminovich et al. 
(\cite{Schiminovich97}),  and they found 
that the neutral hydrogen is all associated to the ring and shows
rotation along this component. 
They estimated a total HI mass of about $2.3 \times 10^9 M_{\odot}$.
This PRG was also mapped by Cox (\cite{Cox96}) in the radio continuum, at 20 cm 
and 6 cm, with the VLA:
she found an extended emission aligned with the ring structure and additional 
filaments which are extended above the ring plane. 
By comparing the radio continuum and the far-infrared (FIR) emission\footnote{The 
far-infrared (FIR) emission was detected $60\mu m$ and
$100\mu m$ by Moshir et al. \cite{Moshir90}, the IRAS Faint-Source
Catalog)}, Cox (\cite{Cox96}) deduced that this PRG falls on the radio/FIR
correlation for star-forming galaxies.

\bigskip

{\bf AM 2020-504} -
Previous photometric and spectroscopic observations 
showed that the central host galaxy in this object is very similar to an 
elliptical galaxy, which is characterized by a decoupled rapidly 
rotating core within $3''$ from the center, (Whitmore et al. 
\cite{whitmore87}; Arnaboldi et al. \cite{magda93}, \cite{magda95}). 
The narrow polar ring, which is observed along the host galaxy minor axis,
is brighter in the optical than in the NIR images (Fig.\ref{hfm1}).  
In all bands, the light distribution of this component is peaked
between $10''$ and $15''$ (see Fig.\ref{prof}). 
As pointed out in the previous Sections,
this object shows different properties with respect to
the other polar ring galaxies studied here. In the high-frequency residual
images (Fig.\ref{hfm1}, middle panel), there is no
trace of any disk-like structures associated with the host galaxy major
axis,  which is observed in all the other polar ring galaxies studied
here. 
The absence of a disk in the host galaxy suggested the use of a
Sersic law (Eq.\ref{sersic}) for the 2D fit of the
light distribution in this component; the results indicate
that the AM 2020-504 central component is an elliptical rather 
than an S0 galaxy.
\bigskip

{\bf ESO 603-G21} -
The prominent structure which appears in the B band image of this object 
is the warped dusty ring (in the SE and NW directions) which
surrounds a bright round stellar system (see Whitmore et al. \cite{PRC} and
Arnaboldi et al. \cite{magda95}). 
This central component is much fainter in the NIR images (Fig.\ref{hfm1}) and 
is embedded in a very luminous disk-like structure. The
high-frequency residual images (Fig.\ref{hfm1}) reveal that this disk is 
nearly edge-on. 
A further fainter ``filamentary'' structure, which was already detected by 
Arnaboldi et al. (\cite{magda95}), is visible perpendicular to this 
disk, and aligned with the apparent major axis of the central 
spheroid ($P.A.=24^o$). The extension of this filamentary structure
is less than $10''$.
The reddest regions in the NIR color maps 
(Fig.\ref{mappe}) correspond to the disk component, while 
the central spheroid is bluer. ESO~603-G21 is the object in our sample
with the reddest nuclear regions (see Table~\ref{polycol}). 
Very recently, Reshetnikov et al. (\cite{Resh02}) have performed a detailed
surface photometry of ESO~603-G21 in the optical B, V and R bands: 
they found that the central component has an exponential light 
distribution and is surrounded by an extended, warped, edge-on disk/ring
structure. 
These results are consistent with our findings.\\
The surface brightness profiles along the bright edge-on disk have an 
exponential behavior (see Fig.\ref{prof}) and the comparison 
with the surface brightness profiles along the orthogonal direction shows that
this disk is the dominant luminous component also in the NIR bands, 
unlike all the other polar rings which are faintest in the NIR. 
This peculiarity makes ESO~603-G26 similar to a late-type spiral galaxy with a 
kinematically-decoupled extended bulge (e.g. NGC~4672 and NGC~4698, see
Bertola et al. \cite{bertola99} and Sarzi et al. \cite{sarzi2000}) rather
than a polar ring galaxy, as also suggested by Reshetnikov et
al. (\cite{Resh02}).  \\
As we will show in Paper~II, the central component of ESO~603-G26 has
colors, age and light distribution properties quite similar to those of
the host galaxy in the other PRGs of our sample. In particular, the
surface brightness distributions is well-fit by the super-position of
two components (bulge + disk) rather than by only an $r^{1/4}$ bulge,
as in NGC~4672 and NGC~4698. The available spectroscopic data for 
this object, which should help us to understand what kind of  
object ESO~603-G26 is,
are quite uncertain (Arnaboldi et al. \cite{magda95}):
they show rotation of the stellar component along the two axis 
corresponding to $P.A.=24^\circ$ and $P.A.=114^\circ$. 
This result may suggest that the underlying central spheroid is triaxial. 
The rotation curve derived from the strong $H\alpha$ emission seen in the
disk/ring spectrum shows a constant velocity gradient along this
component (Arnaboldi et al. \cite{magda95}). This may suggest that we are
indeed looking at a ring.
However, these data are strongly influenced by the dust absorption, 
so no definite conclusion can be derived.\\
The J, H and Ks total magnitudes derived from the 2MASS data, available
for this system, are on average 0.16 mag brighter than the J, H and Kn CASPIR 
magnitudes (see Tab.\ref{totm}). As already discussed in Sec.\ref{totm},
the non-Poissonian background fluctuations, which affects both CASPIR
and 2MASS images, particularly in H band, can explain such differences.\\
Radio data for this object shows CO emission corresponding to 
$1.1 \times 10^9 M_{\odot}$ of molecular hydrogen (Galletta et al. 
\cite{galletta97}) and $6.2 \times 10^9 M_{\odot}$ in HI (van Driel et al. 
\cite{vdriel2000}).
Radio continuum emission was detected for this object by Cox
(\cite{Cox96}) in the central regions of the candidate polar ring 
(i.e. along $P.A.=114^o$, as in the 
PRC). FIR emission, at $60\mu m$ and $100\mu m$, was detected for this PRG 
(Moshir et al. \cite{Moshir90}, the IRAS Faint-Source Catalog),  
this object too falls on the radio/FIR correlation 
for star-forming galaxies rather than for an AGN (Cox \cite{Cox96}).

\section{Summary}\label{conclu}
Here we have presented accurate NIR photometry in J, H and Kn bands for a 
sample of PRGs.
These data were used to compute luminosity profiles, perform bulge-to-disk
decomposition and derive the structural parameters for the central host galaxy
of each PRG in the sample.
Total magnitudes and integrated colors were obtained for the two components
(host galaxy and ring) in all objects. 
A detailed morphological inspection performed for all PRGs in our sample
has revealed the presence of a disk-like structure along the major axis of
the host galaxy in all PRGs of our sample, except for AM~2020-504. This
feature makes the central galaxy similar to an S0 system. In the J-H and
H-K color maps this disk-like structure is quite red. 
The structure-less appearance of AM~2020-504 suggests that 
this object is more similar to an elliptical galaxy.
The extension of the polar structure in all PRGs, except for A0136-0801,
is comparable to the host galaxy optical radius. In the case of A0136-0801
this component is at least 2 times more extended then the host galaxy.
In all PRGs, except for ESO~603-G26, most of the NIR light comes from 
the central host galaxy; in ESO~603-G26 a bright exponential disk 
along the ring direction is the dominant luminous component in these bands.

The NIR photometry obtained for this sample of PRGs is crucial for the
ongoing research about the formation scenarios for these peculiar objects, which
will be discussed in detail in Iodice et al. \cite{paperII} (Paper~II).

\begin{acknowledgements}
We thank the referee, V. Reshetnikov, for several helpful comments and 
suggestions.
E.I and M.A. wish to thank Prof. Capaccioli and the staff from the 
Observatory of Capodimonte for the help and support during the realization
of this work. E.I. would like to thank Prof. F. Combes and Prof. G. Galletta 
for comments and suggestions during her Ph.D. thesis, which is the basis of 
this work.
E.I wish to thank Prof. P. Salucci for the many useful discussions. 
E.I. ad M.A. wish to thank Dr. P. McGregor for providing many information on 
the CASPIR infrared camera.
This publication makes use of data products from the Two Micron All Sky 
Survey (2MASS), which is a joint project of the University of Massachusetts 
and the Infrared Processing 
and Analysis Center/California Institute of Technology, funded by the National 
Aeronautics and Space Administration and the National Science Foundation.
This research has made use of the NASA/IPAC Extragalactic Database (NED) which 
is operated by the Jet Propulsion Laboratory, California Institute of 
Technology, under contract with the National Aeronautics and Space 
Administration.
\end{acknowledgements}

\appendix
\section{Tests on the background noise distribution and error estimate}
\label{error}
\begin{figure}
\centering
\includegraphics[width=6cm]{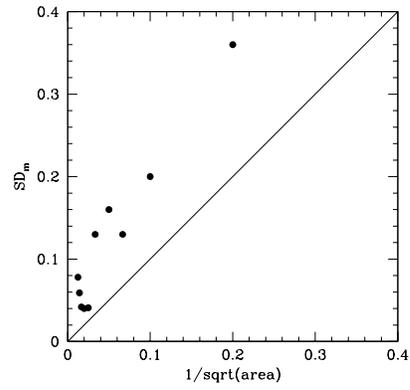}
\caption{Plot of the $SD_m$ vs $1/\sqrt{area}$.
The sloping line corresponds to Poisson noise, if the count in each
pixel has unit standard deviation. The last two
points deviate much more than the others because on larger areas are inevitably 
included bright sources. }
\label{test}
\end{figure}

The uncertainties which affect the integrated magnitudes, derived for each 
object of the sample, are strictly related to the noise in the image 
background level. The {\it background noise} is made up by the residual random 
fluctuations in the data, which appear after the background level has been 
subtracted from the image.
Usually, a Poisson distribution is assumed to describe these fluctuations.
A non correct flatfield correction may introduce further fluctuations in the
background, which may strongly influenced the uncertainties on magnitudes.
Thus, the error estimate, must take into account both the
{\it background noise} and {\it flatfield errors} (see also Matthews and Gallagher 
\cite{MG97}).
In order to verify whether the background noise in the NIR images is well 
described by the Poisson distribution, we sampled the
sky background around the edge of the image, in each band and for each image.
We then performed the following test: in our summed, sky-subtracted images,
we considered boxes on the sky, whose areas increase from 25 $pixels^2$ to even 
larger areas. The largest area depends on how extended the galaxy is on detector field.
For each size, we examined 6 different boxes, in which the mean 
value is derived. We computed the average of
the mean values and derived the relative standard deviation ($SD_m(A)$).
In Fig.\ref{test} is plotted the the $SD_m$ vs $1/\sqrt{area}$, 
derived for the Kn band images.
The diagonal line corresponds to Poisson noise, if the count in each
pixel has unit standard deviation.
This plot suggests that $SD_m$ decreases
as the area increases, but it has larger values than those predicted from
a Poissonian statistics. 
Therefore there is more noise within moderate sky regions
than predicted by a pure counting model. The error estimate must take
into account this ``extra noise".

We estimated the total error in a given area ($A$), which
includes both the Poissonian error and the extra-noise, from
the standard deviation of the mean counts per pixel multiplied by
the area, i.e. $SD_m \times A$.
Furthermore, one should take into account the error due to 
bad flatfield corrections. We can derive an estimate of this quantity with
this second test: we derived the statistics on several (30) sky boxes
(20x20 pixels) for the J, H and Kn images, in which we tried to avoid stars. 
We derived, for each band, the average of the mean values and the standard
deviation ($SD_m$).  These $SD_m$ values give an estimate for the flatfield
error in each band.
The {\it total error}, relative to each pixel value, includes both contributions.
We derived the errors on magnitudes and colors for each bands in the
following way: the integrated magnitudes (Table~\ref{polycol})
are given by $m=ms+q+k-A_\lambda$,
with $ms$ is the instrumental magnitude, $q$ is
the zero point, $k$ and $A_\lambda$ are the correction coefficients for the
atmospheric and galactic extinction. The uncertainties on these last two
parameters are negligible and we consider only the errors relative to 
$ms$ and $q$.
These are independent quantities, so the error in our derived magnitude is 
given by
\begin{equation}
\sigma _m=\sqrt{err_{ms}^2 + err_{q}^2}
\end{equation}
 
with $err_{ms} = 1.0857*(SD_m \times A) flux(A)^{-1}$ and $flux(A)$ is the 
total flux in the $A$ area. 
The average errors on the integrated magnitudes are about
0.05 mag, and an average error of about $13\%$ on NIR colors.
 
There is no easily explanation for the no longer Poissonian distribution of the
background noise. In the NIR images the background level is very high and 
subject to rapid changes: therefore even if carefull subtraction is performed
residuals may still cause deviations from poissonian behaviour.

\end{document}